\documentclass[letter,longauth]{aa}     

\usepackage{graphicx}
\usepackage{txfonts}
\usepackage{lipsum}
\usepackage{subcaption}         
\usepackage{lscape}             
\usepackage{placeins}           

\usepackage{multirow}
\usepackage{makecell}

\usepackage[breaklinks=true]{hyperref}
\usepackage{natbib,twoopt}
\bibpunct{(}{)}{;}{a}{}{,} 

\bibpunct{(}{)}{;}{a}{}{,}             
\makeatletter
  \newcommandtwoopt{\citeads}[3][][]{\href{http://adsabs.harvard.edu/abs/#3}%
    {\def\hyper@linkstart##1##2{}%
     \let\hyper@linkend\@empty\citealp[#1][#2]{#3}}}
  \newcommandtwoopt{\citepads}[3][][]{\href{http://adsabs.harvard.edu/abs/#3}%
    {\def\hyper@linkstart##1##2{}%
     \let\hyper@linkend\@empty\citep[#1][#2]{#3}}}
  \newcommandtwoopt{\citetads}[3][][]{\href{http://adsabs.harvard.edu/abs/#3}%
    {\def\hyper@linkstart##1##2{}%
     \let\hyper@linkend\@empty\citet[#1][#2]{#3}}}
  \newcommandtwoopt{\citeyearads}[3][][]%
    {\href{http://adsabs.harvard.edu/abs/#3}
    {\def\hyper@linkstart##1##2{}%
     \let\hyper@linkend\@empty\citeyear[#1][#2]{#3}}}
\makeatother
\usepackage{graphicx}
\usepackage{booktabs}
\usepackage{lipsum}
\usepackage{txfonts}
\usepackage{hyperref}
\usepackage{tabularx}
\usepackage{upgreek}

\begin{document}

   \title{XRISM Resolves the Circum-nuclear Environment of NGC\,4945}


%

\authorrunning{Boorman et al.}
\titlerunning{XRISM Resolves the Circum-nuclear Environment of NGC\,4945}

\author{Peter~G.~Boorman\inst{\ref{MPE},\ref{Caltech},\ref{Soton}}\thanks{boorman@mpe.mpg.de}
\and Poshak~Gandhi\inst{\ref{Soton}}
\and Yoshihiro~Ueda\inst{\ref{Kyoto}}
\and Bert~Vander~Meulen\inst{\ref{ESTEC}}
\and Kanta~Fujiwara\inst{\ref{Kyoto}}
\and Aoi~Nakano\inst{\ref{Kyoto}}
\and Johannes~Buchner\inst{\ref{MPE}}
\and Claudio~Ricci\inst{\ref{Geneva},\ref{Kavli}}
\and Kirpal~Nandra\inst{\ref{MPE}}
\and Daniel~Stern\inst{\ref{JPL}}
\and Almudena~Alonso~Herrero\inst{\ref{CSICINTA}}
\and Carolina~Andonie\inst{\ref{MPE}}
\and Franz~E.~Bauer\inst{\ref{Tarapaca}}
\and Stefano~Bianchi\inst{\ref{RomaTre}}
\and Murray~Brightman\inst{\ref{Caltech}}
\and Francesca~Civano\inst{\ref{Goddard660}}
\and Richard~I.~Davies\inst{\ref{MPE}}
\and Gulab~Dewangan\inst{\ref{IUCAA}}
\and Georgios~Dimopoulos\inst{\ref{Goddard662}}
\and Ismael~Garc\'{i}a-Bernete\inst{\ref{CSICINTA}}
\and Santiago~Garc\'{i}a-Burillo\inst{\ref{Madrid}}
\and Sebastian~F.~H\"{o}nig\inst{\ref{Soton}}
\and Nancy~A.~Levenson\inst{\ref{STScI}}
\and Andrea~Merloni\inst{\ref{MPE}}
\and Yuya~Nakatani\inst{\ref{Kyoto}}
\and Hirofumi~Noda\inst{\ref{Tohoku}}
\and Shoji~Ogawa\inst{\ref{Yamazaki1},\ref{Yamazaki2}}
\and Chris~Packham\inst{\ref{SanAntonio},\ref{NINS}}
\and Simonetta~Puccetti\inst{\ref{ASI}}
\and Cristina~Ramos~Almeida\inst{\ref{Canarias},\ref{Laguna}}
\and Marko~Stalevski\inst{\ref{Belgrade},\ref{Gent}}
\and Yuichi~Terashima\inst{\ref{Ehime}}
\and Martin~Ward\inst{\ref{Durham}}
\and Satoshi~Yamada\inst{\ref{Tohoku2},\ref{Geneva},\ref{Tohoku}}
}

\institute{
Max Planck Institute for Extraterrestrial Physics, Giessenbachstrasse, 85741 Garching, Germany\label{MPE}
\and
Cahill Center for Astrophysics, California Institute of Technology, 1216 East California Boulevard, Pasadena, CA 91125, USA\label{Caltech}
\and
Department of Physics \& Astronomy, Faculty of Physical Sciences and Engineering, University of Southampton, Southampton, SO17 1BJ, UK\label{Soton}
\and
Department of Astronomy, Kyoto University, Kyoto 606-8502, Japan\label{Kyoto}
\and
European Space Agency (ESA), European Space Research and Technology Centre (ESTEC), Keplerlaan 1, 2201 AZ Noordwijk, The Netherlands\label{ESTEC}
\and
Department of Astronomy, University of Geneva, ch. d’Ecogia 16, 1290 Versoix, Switzerland\label{Geneva}
\and
Kavli Institute for Astronomy and Astrophysics, Peking University, Beijing 100871, People’s Republic of China\label{Kavli}
\and
Jet Propulsion Laboratory, California Institute of Technology, Pasadena, CA 91109, USA\label{JPL}
\and
Centro de Astrobiolog\'{i}a (CAB), CSIC-INTA, Camino Bajo del Castillo s/n, E–28692 Villanueva de la Ca\~{n}ada, Madrid, Spain\label{CSICINTA}
\and
Instituto de Alta Investigaci{\'{o}}n, Universidad de Tarapac{\'{a}}, Casilla 7D, Arica, 1010069, Chile\label{Tarapaca}
\and
Dipartimento di Matematica e Fisica, Universit\`{a} degli Studi Roma Tre, Via della Vasca Navale 84, I-00146, Roma, Italy\label{RomaTre}
\and
NASA Goddard Space Flight Center, Code 660, Greenbelt, MD 20771, USA\label{Goddard660}
\and
Inter-University Centre for Astronomy \& Astrophysics (IUCAA), Pune, 411007, India\label{IUCAA}
\and
NASA Goddard Space Flight Center, Code 662, Greenbelt, MD 20771, USA\label{Goddard662}
\and
Observatorio Astron\'{o}mico Nacional (OAN–IGN), Observatorio de Madrid, Alfonso XII, 3, E-28014 Madrid, Spain\label{Madrid}
\and
Space Telescope Science Institute, 3700 San Martin Drive, Baltimore, MD 21218, USA\label{STScI}
\and
Astronomical Institute, Tohoku University, 6-3 Aramakiazaaoba, Aoba-ku, Sendai, Miyagi 980-8578, Japan\label{Tohoku}
\and
Faculty of Science and Technology, Tokyo University of Science, 2641 Yamazaki, Noda-shi, Chiba 278-8510, Japan\label{Yamazaki1}
\and
Research Center for Space System Innovation, Research Institute for Science \& Technology, Tokyo University of Science, 2641 Yamazaki, Noda-shi, Chiba 278-8510, Japan\label{Yamazaki2}
\and
Department of Physics and Astronomy, University of Texas at San Antonio, One UTSA Circle, San Antonio, TX 78249, USA\label{SanAntonio}
\and
National Astronomical Observatory of Japan, National Institutes of Natural Sciences (NINS), 2-21-1 Osawa, Mitaka, Tokyo 181-8588, Japan\label{NINS}
\and
ASI-Agenzia Spaziale Italiana, Via del Politecnico snc, 00133, Rome, Italy\label{ASI}
\and
Instituto de Astrofísica de Canarias, Calle V\'{i}a L\'{a}ctea, s/n, 38205 La Laguna, Tenerife, Spain\label{Canarias}
\and
Departamento de Astrof\'{i}sica, Universidad de La Laguna, 38206 La Laguna, Tenerife, Spain\label{Laguna}
\and
Astronomical Observatory, Volgina 7, 11060 Belgrade, Serbia\label{Belgrade}
\and
Department of Physics and Astronomy, Universiteit Gent, Proeftuinstraat 86 N3, B-9000 Ghent, Belgium\label{Gent}
\and
Department of Physics, Ehime University, 2-5 Bunkyo-cho, Matsuyama, Ehime 790-8577, Japan\label{Ehime}
\and
Centre for Extragalactic Astronomy, Department of Physics, Durham University, South Road, Durham DH1 3LE, UK\label{Durham}
\and
Frontier Research Institute for Interdisciplinary Sciences, Tohoku University, Sendai, Miyagi 980-8578, Japan\label{Tohoku2}\\
}

   \date{}

 
  \abstract{Compton-thick Active Galactic Nuclei (AGN) represent one of the most elusive phases of massive black hole growth, yet are expected to contribute substantially to the Cosmic X-ray Background and the integrated growth of massive black holes. NGC\,4945 is the closest Compton-thick AGN and amongst the brightest AGN in the hard X-ray sky, making it an important benchmark for more distant Compton-thick AGN. We present the first high-resolution X-ray spectral analysis of NGC\,4945 using \textit{XRISM}/Resolve. The entire 4\,--\,15\,keV Resolve spectrum, including a strong Fe\,K$\alpha$ doublet and weak Compton Shoulder, is well described by a de-coupled dual-obscurer model. The model features a low-covering-factor Compton-thick primary obscurer intersecting the line-of-sight that permits the rapidly variable, direct transmitted coronal continuum to dominate above 10\,keV. A Compton-thin secondary reprocessor with a high covering factor dominates the reprocessed emission between $\sim$\,4\,--\,10\,keV. Assuming that virial motion accounts for line broadening, the secondary reprocessor can exist at $\sim$\,0.12\,pc, and could help explain the weak high-ionisation optical and infrared emission lines observed in NGC\,4945. If such obscuration geometries are common among more distant and/or fainter Compton-thick AGN, our results suggest that simpler coupled X-ray spectral modelling could substantially over-estimate Compton-thick covering factors and under-estimate intrinsic X-ray luminosities.}

   \keywords{galaxies: active --
                galaxies: individual: NGC\,4945 --
                X-rays: galaxies
               }

   \maketitle
\nolinenumbers

\section{Introduction}

Recent X-ray stacking of high-redshift Active Galactic Nuclei (AGN) identified by \textit{JWST} \citep{Maiolino25,Comastri26} has reinforced predictions from X-ray surveys that a significant fraction of supermassive black hole growth across cosmic time occurred beneath Compton-thick obscuration \citep{Ueda14,Ricci15,Ananna19,Boorman25_nulands}. However, the Compton-thick AGN population remains one of the most elusive and poorly-characterised AGN populations at any redshift. The first Data Release of the Database of Compton-Thick AGN \citep{Boorman24_hexp} identified just 66 local AGN at $\lesssim$\,400\,Mpc with at least one published \textit{NuSTAR} spectral fit yielding a Compton-thick column density, $N_{\rm H}$\,$>$\,1.5\,$\times$\,10$^{24}$\,cm$^{-2}$, to $\geq$\,90\% confidence. In addition, the hard X-ray signal-to-noise ratio distribution of all 66 Compton-thick AGN was dominated by three nearby targets: NGC\,4945, the Circinus Galaxy, and NGC\,1068. Consequently, these three sources are essential for obtaining the detailed insights needed to interpret high-redshift Compton-thick AGN.

NGC\,4945 is the closest of all known Compton-thick AGN \citep{Iwasawa93} and the third brightest radio-quiet AGN in hard X-rays from the 105-month \textit{Swift}/BAT catalogue (observed flux $F_{14-195\,{\rm keV}}$\,$\sim$\,3\,$\times$\,10$^{-10}$\,erg\,s$^{-1}$\,cm$^{-2}$; \citealt{Oh18}, see also \citealt{Done96}). It lies in a nearby (3.7\,Mpc; \citealt{Tully15}) edge-on spiral galaxy with a highly obscured composite nuclear starburst/AGN \citep{Moorwood94} with a disc-megamaser-based black hole mass of $\sim$\,$1.4 \times 10^6$\,M$_{\odot}$ \citep{Greenhill97}. NGC\,4945 varies strongly above 10\,keV \citep{Madejski00,Yaqoob12,Puccetti14}, spanning flux factor changes of $\sim$\,1.5\,--\,2 within a day, with luminosities peaking at $\gtrsim$\,30\% of its Eddington limit \citep{Done03,Puccetti14}. Such strong variability suggests that the direct transmitted coronal emission escapes at $\gtrsim$\,10\,keV, and many previous X-ray spectral models have required very small covering factors, of order $\lesssim$\,20\%, for the Compton-thick obscurer to explain it (e.g., \citealt{Madejski00,Itoh08,Yaqoob12,Puccetti14}). However, \citet{Brightman15} showed that very high covering factor models can still explain the observed \textit{NuSTAR} data and are in agreement with the luminosity-dependent scaling of covering factors measured for other local Compton-thick AGN. \textit{Chandra} has spatially resolved part of the obscurer at $\sim$\,30\,pc scales \citep{Marinucci12,Marinucci17}, suggesting multiple obscurers are responsible for the observed X-ray properties of NGC\,4945 (see also \citealt{Done03}). The edge-on host galaxy complicates segregation of circum-nuclear vs. host galaxy obscuration, and may contribute to its apparent mid-infrared under-luminous position on the infrared vs. X-ray luminosity correlation \citep{Goulding2012,Gandhi15_n4785,Asmus2015}.

The Resolve microcalorimeter \citep{Ishisaki25} onboard \textit{XRISM} \citep{Tashiro25} is providing new insights into the structure and composition of circum-nuclear obscurers surrounding Compton-thick AGN. For example, Resolve has revealed differing Fe\,K$\alpha$ Compton Shoulder profiles from the Circinus Galaxy \citep{XRISMCollaboration26} and NGC\,1068 \citep{Bianchi26}, despite expectations for similar profiles to arise from Compton-thick reprocessing (e.g., \citealt{Matt02,Dimopoulos24}). Here we present the first detailed high-resolution X-ray spectral analysis of NGC\,4945 using Resolve, with the aim of understanding the high-spectral-resolution X-ray signatures of its circum-nuclear obscurer(s).

\section{Data and Methodology}\label{sec:data}

\textit{XRISM} observed NGC\,4945 on 2025--June--23 for a cleaned exposure time of 291\,ks (Obs.\,ID 201094010, PI: P.~G.~Boorman). In this paper, we focus on the Resolve data and defer analysis of the Xtend data to future work. Unfiltered event files were reprocessed with \texttt{xapipeline} using \texttt{CALDB v\,gen20241115\_xtd20241115\_rsl20241115} within \texttt{HEASoft v6.35}. Cleaned level 2 event files were produced by screening out pixel-pixel coincident events, removing anomalous Ls events, only selecting High-primary events and excluding pixel 27 data as described in the \textit{XRISM} Data Reduction guide\footnote{\url{https://heasarc.gsfc.nasa.gov/docs/xrism/analysis/abc_guide/xrism_abc.html}}. \texttt{XSELECT} was used to extract spectra and light curves, and the \texttt{rslmkrmf} command was used to generate response matrix files. Exposure maps were created with the \texttt{xaexpmap} command before constructing ancillary response files with the \texttt{xaarfgen} command using 300,000 simulated photons. To investigate the level of non-X-ray-background present in the extracted spectrum, we used the \texttt{rslnxbgen} command following the Spectral Extraction Recipes\footnote{\url{https://heasarc.gsfc.nasa.gov/docs/xrism/analysis/nxb/resolve_nxb_db.html}}. The non-X-ray-background was found to be sub-dominant to the source emission across the 4\,--\,15\,keV passband, and so is neglected hereafter.

We fit the un-subtracted source\,$+$\,background time-averaged spectrum over the 4\,--\,15\,keV passband with \texttt{PyXspec v12.12.1} \citep{Arnaud96,Gordon21}, using the Poisson Likelihood (\texttt{cstat}; \citealt{Cash79}). Our justification and assumptions associated with analysing the time-averaged spectrum, as opposed to spectra separated into distinct flux levels, are provided in §\ref{sec:timeres}. Though no binning is required with \texttt{cstat} \citep{Buchner23a}, we bin the spectrum using the `optimal' binning scheme of \citet{Kaastra16} to reduce the computation time of parameter estimation without a loss of information. We use the Bayesian X-ray Analysis package (\texttt{BXA}; \citealt{Buchner14,Buchner21}) \texttt{v2.9}, which connects the nested sampling algorithm \texttt{MultiNest} \citep{Feroz09} to \texttt{PyXspec}. All fits were performed with a sampling efficiency of 0.3 and 400 live points as a compromise between a feasible computation time and robust posterior reconstruction \citep{Dittmann24}. All spectral models included Galactic absorption parameterised with \texttt{TBabs} and $N_{\rm H}$\,=\,2.17\,$\times$\,10$^{21}$\,cm$^{-2}$ \citep{Willingale13} using the \citet{Wilms00} abundances. We measure the velocity offset relative to the systemic redshift of 0.002 \citep{Koss22} by allowing the redshift to vary with a log-uniform prior. All parameter values and uncertainties are quoted as the mode and 90\% Highest Density Interval of their respective marginalised posterior distributions.

\section{X-ray Spectral Modelling Results}\label{sec:results}

\begin{figure*}[h!]
    \centering
    \includegraphics[width=\textwidth]{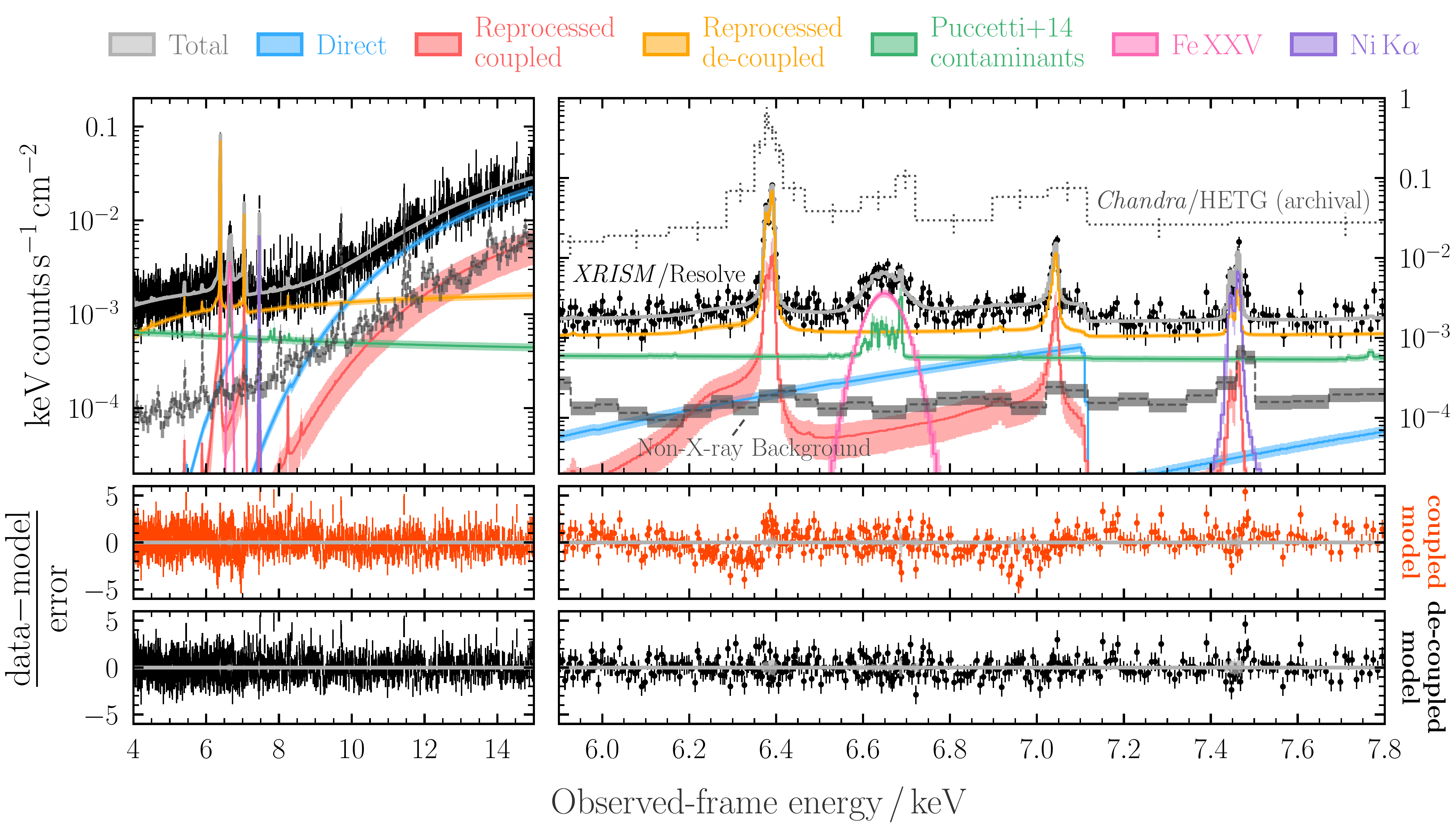}
    \caption{(\textit{Upper panels}) Resolve spectral fit with the de-coupled \texttt{XSKIRTOR} model over the 4\,--\,15\,keV and 5.8\,--\,7.8\,keV passbands on the left and right, respectively. The upper right panel also presents the archival 171\,ks \textit{Chandra}/HETG spectrum for comparison. (\textit{Centre panels}) Residuals from the dis-favoured coupled model. (\textit{Lower panels}) Residuals from the favoured de-coupled model.}
    \label{fig:specres}
\end{figure*}

To constrain the X-ray obscurer(s) within NGC\,4945, we use the high-resolution \texttt{XSKIRTOR} model\footnote{\url{https://github.com/BertVdM/xskirtor}} (\citealt{VanderMeulen23}, Vander Meulen, subm.), including the intrinsic emission-line profiles measured by \citet{Holzer97}. The model assumes solar abundances for all elements considered, including Fe and Ni that are particularly pertinent for our analysis. The geometry is a sphere with polar-cone cut-outs, parameterised by the line-of-sight column density, covering factor ($\equiv$\,cos\,$\theta_{\rm OA}$, where $\theta_{\rm OA}$ is the cut-out half-opening angle) and inclination relative to the pole. The X-ray coronal emission is modelled as an exponentially cut-off power law with variable photon index, normalisation and exponential cut-off energy\footnote{Due to the lack of spectral coverage at $>$\,15\,keV, we fix the cut-off energy to 200\,keV \citep{Ricci18,Balokovic20}.}. Compton-thick AGN spectral modelling is model-dependent, particularly for intrinsic luminosity estimates \citep{Kallova24,Boorman24b}; quantifying these systematics is beyond the scope of this study and will be addressed in future work.

We first fit the Resolve spectrum using a coupled configuration similar to \citet{Brightman15}, in which the intrinsic emission is absorbed by the same obscurer responsible for Compton scattering and fluorescence. We fix its inclination to edge-on, assuming the disc megamaser traces the same physical structure (c.f. \citealt{Madejski00,Itoh08}). The four free obscurer/coronal parameters are the photon index, with a Gaussian prior of mean 1.8 and standard deviation 0.05 (c.f. \citealt{Yaqoob12,Puccetti14}); the intrinsic coronal normalisation and line-of-sight column density, both with log-uniform priors; and the obscurer covering factor, with a uniform prior. Given Resolve's enhanced spectral resolution, we also include variable Gaussian smoothing of the reprocessed component and Ni\,K$\alpha$ emission (added to the Ni\,K$\alpha$ emission inherent to \texttt{XSKIRTOR}; c.f. §\ref{subsec:nika}) with log-uniform priors. Together with 18 nuisance parameters (including the treatment of Fe\,XXV; c.f. §\ref{sec:fexxv}) and the variable redshift gives 24 free parameters.

The coupled-model fit parameters are shown in Table~\ref{tab:table1}. The residuals arising from the fit (Fig.~\ref{fig:specres}) show that it cannot simultaneously reproduce the strong Fe\,K$\alpha$ line and weak Compton Shoulder. The line-of-sight column density is Compton-thick with log\,$N_{\rm H,\,LOS}$\,/\,cm$^{-2}$\,=\,$24.384^{+0.005}_{-0.007}$ and the covering factor reaches its maximum, CF$_{1}$\,=\,95$^{+u}_{-3}$\%, consistent with \citet{Brightman15}. Above 10\,keV, the spectrum is also dominated by Compton-thick reprocessed emission, contrary to the interpretation of the strong hard X-ray variability discussed in §\ref{sec:timeres} (see also \citealt{Yaqoob12}). By including the known X-ray contaminants characterised by \citet{Puccetti14}, the coupled model disfavours a solution in which contaminating flux saturates the Compton Shoulder. Thus, the Compton-thick line-of-sight column density and high covering factor appear to be at odds with the weak Compton Shoulder arising from Compton-scattered fluorescence photons. This suggests that additional emission contributes to the observed continuum and lines.

De-coupling is a common alternative for fitting high signal-to-noise ratio X-ray spectra of obscured AGN \citep{Lamassa19,Boorman24_hexp}, allowing the reprocessed flux to be independent of the line-of-sight absorption \citep{Yaqoob12}. To ensure physical self-consistency, we first retain a primary coupled obscurer with the setup described above, then include emission from a secondary reprocessor with an unobstructed view of the corona. Its illuminating photon index and normalisation are tied to the corona, while its column density and covering factor vary independently with log-uniform and uniform priors, respectively. Its line-of-sight absorption is not considered since the secondary reprocessor does not lie along the line-of-sight. The cosine of the secondary inclination varies with a uniform prior, and the secondary reprocessed component has a multiplicative scaling and independent Gaussian smoothing, each with a log-uniform prior. This configuration has 29 free parameters.

The de-coupled model (Fig.~\ref{fig:specres}, Table~\ref{tab:table1}) reproduces the Resolve spectrum well, including the major emission lines and weak Fe\,K$\alpha$ Compton Shoulder. The primary obscurer remains Compton-thick, with log\,$N_{\rm H,\,LOS}$\,/\,cm$^{-2}$\,=\,24.58\,$\pm$\,0.01, but has a lower covering factor, CF$_{1}$\,=\,$10$\,$\pm$\,$5$\%. The secondary reprocessor has a substantially lower column, log\,$N_{\rm H,\,2}$\,/\,cm$^{-2}$\,=\,22.96$^{+0.18}_{-0.13}$, and a larger covering factor, CF$_{2}$\,=\,47$^{+u}_{-18}$\%. Its reprocessed flux dominates at $\sim$\,4\,--\,10\,keV, while direct transmitted emission dominates at $>$\,10\,keV, as required by the hard X-ray variability.

\section{Discussion and Implications}\label{sec:discussion}

\begin{figure*}[h!]
    \centering
    \includegraphics[width=\textwidth]{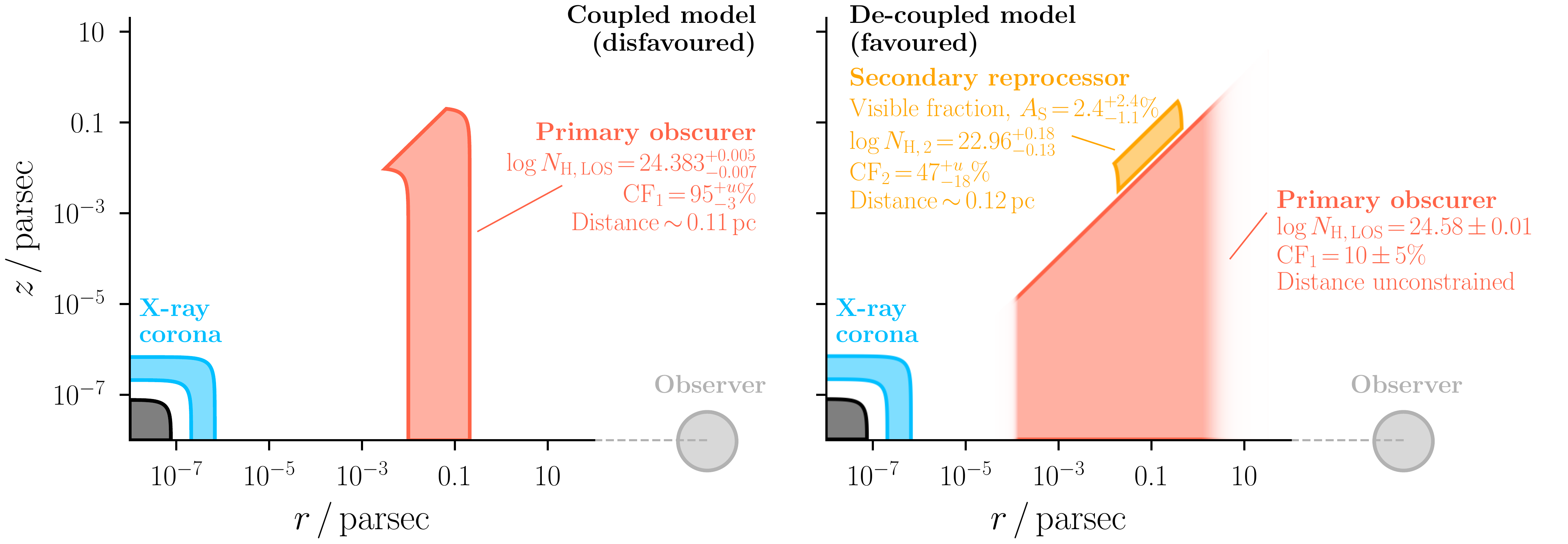}
    \caption{Schematic illustration of the circum-nuclear environment of NGC\,4945 inferred from the coupled (\textit{left}) and de-coupled (\textit{right}) X-ray spectral fits to the Resolve spectrum. All geometries are presented as annular wedges, scaled to log-radial units.}
    \label{fig:schematic}
\end{figure*}

The de-coupled model implies a primary obscurer (assumed to be edge-on), and a secondary reprocessor with covering factors of CF$_{1}$\,=\,$10$\,$\pm$\,$5$\% and CF$_{2}$\,=\,47$^{+u}_{-18}$\%, respectively. Independent Gaussian broadening of each reprocessed spectrum provides virial distance estimates (c.f. \citealt{Gandhi15,Andonie22,Andonie22b}). Although some broadening is generally required for the Compton-thick obscurer, its FWHM is unconstrained towards lower values, allowing solutions in which its lines are only weakly resolved by \textit{XRISM}. Including a 50\% black-hole-mass uncertainty \citep{Brightman16}, its distance is consequently unconstrained over $\sim$\,10$^{-4}$\,--\,30\,pc. However, because \textit{Chandra} imaging has resolved part of the obscurer on $\sim$\,30\,pc scales \citep{Marinucci12}, larger distances are unlikely if the primary obscurer traces the same component. The secondary reprocessor lies at $\sim$\,0.12\,pc from the X-ray corona (99.7\% upper limit of 0.46\,pc) and has an inclination angle of $\theta_{\rm inc,\,2}$\,=\,$71^{+u}_{-22}$$^{\circ}$. This permits the two obscuring mediums to occupy spatially distinct locations. Fig.~\ref{fig:schematic} compares this configuration with the disfavoured coupled-model fit and its corresponding $\sim$\,0.11\,pc distance inferred from Gaussian broadening.

The secondary-to-primary reprocessed-emission scaling is $A_{\rm S}$\,=\,2.4$^{+2.4}_{-1.1}$\%. The de-coupled fit thus requires only a small effective fraction of the secondary reprocessed flux to reach the observer without additional obscuration. As shown in Fig.~\ref{fig:schematic}, the primary obscurer could shield the secondary reprocessor, particularly if it extends to $\sim$\,30\,pc \citep{Marinucci12}. Thus, $A_{\rm S}$ could represent clumpiness in the primary obscurer that allows the required secondary reprocessed flux (possibly originating at the inner edge of the primary obscurer) to escape along unobscured sight lines. Alternatively, $A_{\rm S}$ could reflect partial ionisation of either obscurer, warps in the maser disk assumed to trace the Compton-thick obscurer \citep{Greenhill97} or coronal variability, with the two reprocessed components responding to different time-averaged illuminating fluxes (e.g., \citealt{Yaqoob12,Tanimoto19}). We therefore interpret $A_{\rm S}$ phenomenologically as an effective parameter accounting for several processes not included explicitly in the spectral fits.

The secondary reprocessor's relatively unconstrained inclination and large covering factor are consistent with polar gas \citep{McKaig22}, an extended polar wind \citep{Wada12,Hoenig19} and/or a `hot mirror' funnelled by the primary Compton-thick obscurer \citep{Boorman18,Matt19}. If compact, the secondary reprocessor could suppress the polar mid-infrared extended emission absent from sub-arcsecond ground-based imaging \citep{Asmus2015}\footnote{See also: \url{https://dc.zah.uni-heidelberg.de/sasmirala/q/prod/qp/NGC\%204945}}. This may be consistent with \citet{PerezBeaupuits11}, who argue that a high obscuring covering factor is required to explain the weak [Ne\,V] and [O\,IV] coronal lines observed from NGC\,4945. Weak high-excitation infrared and optical lines have likewise been interpreted as evidence that a Compton-thin obscurer suppresses lower-energy X-rays over many directions \citep{Done03}, as permitted by our de-coupled model. Gas channelled by the ongoing nuclear starburst towards the corona is one plausible origin \citep{Done03}. If the reprocessor is instead a radially extended wind, its virial distance estimate need not apply and it could extend to larger radii. A geometrically thin Compton-thick disc plus a larger-scale, geometrically thick Compton-thin outflow is a state-of-the-art interpretation of the AGN obscuring `torus' \citep{Wada15,Hoenig19,Williamson20}. Furthermore, X-ray reprocessing from a wind could in principle explain the emission lines required for the secondary reprocessor (e.g., \citealt{Matzeu22}), and partial ionisation could produce the required effective secondary flux reduction.

NGC\,4945 is sufficiently nearby and bright to serve as a potential template for other more distant and/or fainter Compton-thick AGN. Our results suggest that applying simpler coupled models to similar systems could yield over-estimated covering factors and under-estimated intrinsic luminosities. The Compton-thick covering factor constrained with our de-coupled model is significantly below the Compton-thick AGN fraction inferred from local isotropic surveys \citep{Boorman25_nulands,Annuar25}, possibly suggesting its high inferred Eddington ratio (c.f. Table~\ref{tab:table1}) suppresses the Compton-thick obscurer through the effective Eddington limit on dusty gas \citep{Fabian08,Ricci17_Nat}. Future spatially resolved, spectro-temporal studies combining simultaneous \textit{XRISM}, \textit{Chandra} and \textit{NuSTAR} observations with obscuration models that incorporate variable ionisation states will be key to constrain the corona and circum-nuclear obscurer of NGC\,4945 in unprecedented detail.

\bibliographystyle{aa}
\bibliography{bib}

\begin{appendix}
\nolinenumbers
\onecolumn




\section{Acknowledgements}
\begin{acknowledgements}
PGB acknowledges support under NASA grant 80NSSC25K7481 and NASA contract NNG08FD60C. PG was a Royal Society Leverhulme Trust Senior Research Fellow during part of this work (SRF\textbackslash R1\textbackslash 241074), and  thanks the Science and Technology Facilities Council (ST/Y001680/1) for support. The work of DS was carried out at the Jet Propulsion Laboratory, California Institute of Technology, under a contract with the National Aeronautics and Space Administration (80NM0018D0004). FEB acknowledges support from ANID-Chile BASAL CATA FB210003 and FONDECYT Regular 1241005. SGB acknowledges support from the Spanish grant PID2022-138560NB-I00, funded by MCIN/AEI/10.13039/501100011033/FEDER, EU. This study was supported by the Japan Society for the Promotion of Science (JSPS) KAKENHI via grant Nos. of 24K00672 and 25H00660. This work was supported by the Japan Society for the Promotion of Science (JSPS) KAKENHI grant number 24K17104 (S.O.). CRA acknowledges support from the Agencia Estatal de Investigaci\'{o}n of the Ministerio de Ciencia, Innovaci\'{o}n y Universidades (MCIU/AEI) under the grant ``Tracking active galactic nuclei feedback from parsec to kiloparsec scales'', with reference PID2022-141105NB-I00 and the European Regional Development Fund (ERDF). MS was supported by the Ministry of Science, Technological Development and Innovation of the Republic of Serbia (MSTDIRS) through contract no. 451-03-33/2026-03/200002 with the Astronomical Observatory (Belgrade). BV acknowledges support through the European Space Agency (ESA) Research Fellowship in Space Science. This research is based on observations obtained with \textit{XRISM}, a JAXA/NASA collaborative mission, with ESA participation. This research has made use of data and/or software provided by the High Energy Astrophysics Science Archive Research Center (HEASARC), which is a service of the Astrophysics Science Division at NASA/GSFC and the High Energy Astrophysics Division of the Smithsonian Astrophysical Observatory. This research has made use of the NASA/IPAC Extragalactic Database (NED), which is operated by the Jet Propulsion Laboratory, California Institute of Technology, under contract with the National Aeronautics and Space Administration.
This research has made use of NASA’s Astrophysics Data System Bibliographic Services. This paper made extensive use of \texttt{matplotlib} \citep{Hunter2007}, \texttt{pandas} \citep{reback2020pandas, mckinney-proc-scipy-2010} and \texttt{astropy} \citep{astropy:2013,astropy:2018,astropy:2022}.
\end{acknowledgements}

\section{Time-resolved Analysis}\label{sec:timeres}
\begin{figure*}[h!]
    \centering
    \includegraphics[width=\textwidth]{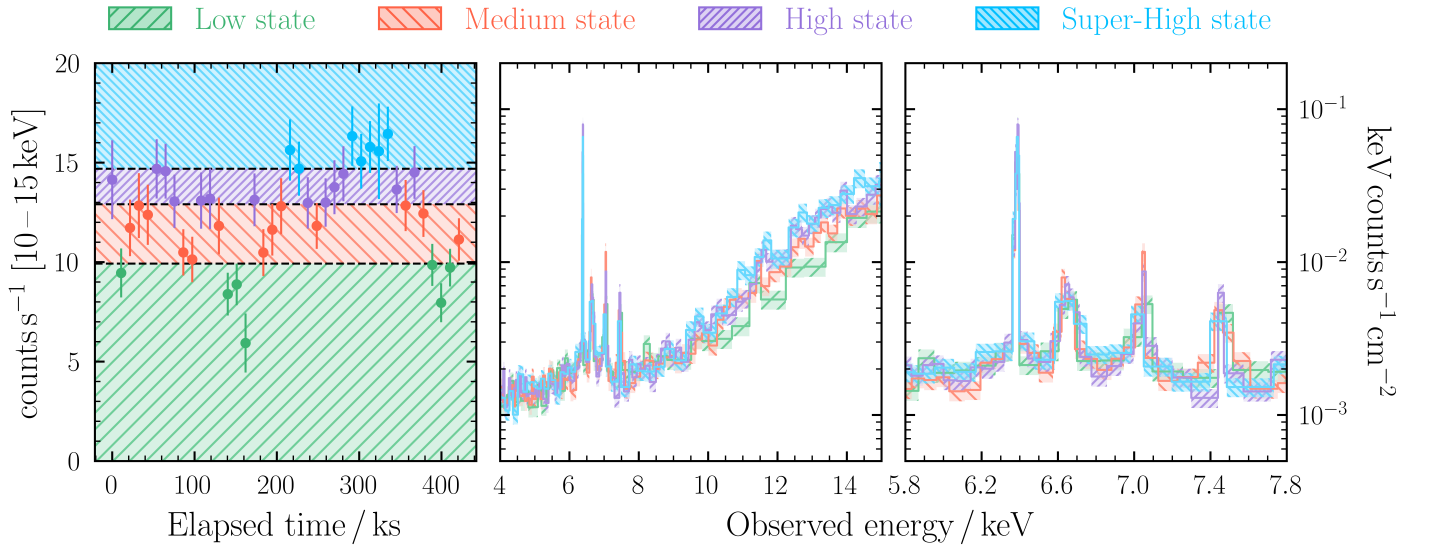}
    \caption{(\textit{Left}) \textit{XRISM}/Resolve light curve in 10\,--\,15\,keV, separated into four distinct flux states. (\textit{Center}) Corresponding 4\,--\,15\,keV Resolve spectra extracted for each of the four flux states shown in the left panel. Clear variability is seen at $>$\,10\,keV. (\textit{Right}) A zoom-in of the center panel in the 5.8\,--\,7.8\,keV range, highlighting a lack of significant variability in any emission lines or underlying Compton-scattered continuum.}
    \label{fig:timeres}
\end{figure*}

The left panel of Fig.~\ref{fig:timeres} presents the 10\,--\,15\,keV \textit{XRISM}/Resolve light curve of the source, showing clear hard X-ray variability. We thus flux-resolved the light curve into four states quantified as the 16th, 50th and 84th count rate quantiles. The centre panel of Fig.~\ref{fig:timeres} presents the 4\,--\,15\,keV spectra separated into these distinct flux levels, in which the significant variability at $\gtrsim$\,10\,keV is evident. The right panel of Fig.~\ref{fig:timeres} shows there is no significant variability detected over the 5.8\,--\,7.8\,keV passband throughout the observation. Due to the computation time of simultaneously fitting multiple flux-resolved spectra, we hereafter focus on constructing a spectral model to explain the assumed non-variable obscurer(s) of NGC\,4945 and average intrinsic emission assumed to dominate at $\gtrsim$\,10\,keV. Changes in the intrinsic coronal spectral shape may also contribute (e.g., \citealt{Puccetti14}), but are not tested in the present analysis. As such, all spectral fits in this work were performed over the 4\,--\,15\,keV passband. Our choice to include energies above the nominal Resolve maximum energy of 12\,keV was primarily driven by our ability to constrain the intrinsic coronal emission emerging at $>$\,10\,keV.

\section{Spectral Model Nuisance Parameters}\label{sec:nuisancepars}

\subsection{Off-nuclear contaminants}\label{sec:offnuc}
NGC\,4945 hosts a nuclear starburst and wide range of X-ray emitting off-nuclear components that contaminate the AGN X-ray signatures at $\lesssim$\,10\,keV \citep{Puccetti14,Brightman23} that are inadvertently included in our \textit{XRISM}/Resolve spectrum of the source. As a means to account for possible flux contamination in our X-ray spectral fits, we include bespoke priors to marginalise over the contaminants with nuisance parameters. \citet{Puccetti14} presented a detailed spatially-resolved analysis of the contaminating sources by using archival \textit{Chandra}/ACIS data to resolve structures that were not spatially resolved by \textit{Suzaku} within their 1.85\,arcmin extraction radius. This circular aperture is similar to the 3.1\,arcmin$^{2}$ field-of-view of \textit{XRISM}/Resolve that we extract the spectra from, meaning the level of contamination found with \textit{Suzaku} should be comparable to our case with Resolve. We note that NGC\,4945 is also known to host transient X-ray sources (e.g., \citealt{Ide20,Brightman23}), that may by chance be present during the \textit{XRISM} observation of NGC\,4945 but not during the \textit{Chandra} observations, or vice versa. However, for the present work we assume that any bright X-ray transients are rare and comparatively weak enough to neglect their contribution in our overall spectral modelling of the AGN.

We fully incorporate the prior information regarding spectral contaminants in NGC\,4945 identified by \citet{Puccetti14} into our spectral modelling. This was achieved by including nine additional model components featuring 14 free parameters that followed the exact constraints given in Tables~5 and~6 as well as §5.2.3 from \citet{Puccetti14} assuming an asymmetric Gaussian distribution for all priors. We note that their use of \texttt{mekal} and \texttt{wabs} was updated to \texttt{apec} and \texttt{TBabs} respectively, once we verified there were no substantial changes in overall flux between both sets of components.

\subsection{Fe\,XXV}\label{sec:fexxv}
The extracted \textit{XRISM}/Resolve spectrum of NGC\,4945 shown in Fig.~\ref{fig:specres} clearly reveals strong, seemingly broad, emission at a rest-frame energies of $\sim$\,6.67\,keV coincident with Fe\,\textsc{XXV}. \citet{Bianchi26} recently explained the complex Fe\,\textsc{XXV} and Fe\,\textsc{XXVI} emission of NGC\,1068 with a biconical outflow model fundamentally powered by the AGN. However, for NGC\,4945 it is not obvious if the origin of the Fe\,\textsc{XXV} emission should be considered AGN-driven, powered by the circum-nuclear starburst or a combination of both. In this paper we choose to treat the Fe\,\textsc{XXV} emission from NGC\,4945 as a nuisance parameter and defer its detailed spectroscopic treatment to a future work. We sought to reproduce the Fe\,\textsc{XXV} emission with as simplistic a phenomenological prescription as possible. After multiple trials involving multiple Gaussian lines, it was found that a single broad Gaussian emission line reproduced the feature satisfactorily. Thus all X-ray spectral modelling included an additional single broad Gaussian line with three variable parameters: centroid energy, line width and normalisation.

\subsection{Ni\,K$\alpha$}\label{subsec:nika}
Preliminary tests revealed that standard solar levels of Nickel were insufficient to explain the Ni\,K$\alpha$ emission detected by \textit{XRISM}/Resolve for NGC\,4945. Due to the overall strength of the Ni\,K$\alpha$ emission line detected, we included additional Ni\,K$\alpha$ emission as a nuisance parameter to avoid any unnecessary biases to the overall estimate of the continuum. To do this, we included the intrinsic line profile of Ni\,K$\alpha$ as parameterised by \citet{Holzer97} with the summation of five Lorentzian functions (identical to the implementation in \texttt{XSKIRTOR}; \citealt{VanderMeulen24_lines}). The additional Ni\,K$\alpha$ emission thus contributed one unique additional free parameter to all corresponding spectral fits: the overall normalisation of the line profile. The resulting Nickel abundances quoted in Table~\ref{tab:table1} are computed via a separate local fit to the Ni\,K$\alpha$ doublet using the intrinsic line profile of \citet{Holzer97} and the resulting additional Ni\,K$\alpha$ emission required in the coupled and de-coupled model fits. Fully-consistent super-Solar Nickel abundances were required in either fits to NGC\,4945, comparably to the recent \textit{XRISM} findings for the Circinus Galaxy and NGC\,1068 \citep{XRISMCollaboration26,Bianchi26}.

\section{Spectral Model Parameter Constraints}\label{sec:parres}
Table~\ref{tab:table1} presents the parameter values constrained from the coupled and de-coupled X-ray spectral model fits.

\renewcommand{\arraystretch}{1.4}
\begin{table}
\centering
\caption{Spectral parameter results for both models tested.}
\label{tab:table1}
\begin{tabular*}{\textwidth}{rrccl}
\toprule
                                                            Component &                          Parameter &                    Coupled &              De-coupled &                                                  Unit \\
\midrule
                                    \multirow{4}{*}{Coronal emission} &                     $\Gamma$$^{a}$ &      $1.70$\,$\pm$\,$0.04$ &   $1.72$\,$\pm$\,$0.05$ &                                                    -- \\
                                                                      &                E$_{\rm cut}$$^{b}$ &                  200$^{*}$ &               200$^{*}$ &                                                   keV \\
                                                                      &   log\,$L_{2-10\,{\rm keV}}$$^{c}$ &     $41.52$\,$\pm$\,$0.03$ &  $42.64$\,$\pm$\,$0.11$ &                                         erg\,s$^{-1}$ \\
                                                                      &          $\lambda_{\rm Edd}$$^{d}$ &            $5$\,$\pm$\,$2$ &        $54^{+35}_{-29}$ &                                                    \% \\
\midrule
                                    \multirow{4}{*}{Primary obscurer} &       log\,$N_{\rm H,\,LOS}$$^{e}$ & $24.383^{+0.005}_{-0.007}$ &  $24.58$\,$\pm$\,$0.01$ &                                             cm$^{-2}$ \\
                                                                      &                     CF$_{1}$$^{f}$ &             $95^{+u}_{-3}$ &        $10$\,$\pm$\,$5$ &                                                    \% \\
                                                                      &       $\theta_{\rm inc,\,1}$$^{g}$ &                   90$^{*}$ &                90$^{*}$ &                                            $^{\circ}$ \\
                                                                      &           FWHM$_{1}$(6\,keV)$^{h}$ &          $265^{+27}_{-31}$ &    $387^{+196}_{-u}$ &                                          km\,s$^{-1}$ \\
\midrule
                                  \multirow{5}{*}{Secondary reprocessor} &                  $A_{\rm S}$$^{i}$ &                         -- &     $2.4^{+2.4}_{-1.1}$ &                                                    \% \\
                                                                      &         log\,$N_{\rm H,\,2}$$^{j}$ &                         -- & $22.96^{+0.18}_{-0.13}$ &                                             cm$^{-2}$ \\
                                                                      &                     CF$_{2}$$^{k}$ &                         -- &        $47^{+u}_{-18}$ &                                                    \% \\
                                                                      &       $\theta_{\rm inc,\,2}$$^{l}$ &                         -- &        $71^{+u}_{-22}$ &                                            $^{\circ}$ \\
                                                                      &           FWHM$_{2}$(6\,keV)$^{m}$ &                         -- &       $261^{+44}_{-77}$ &                                          km\,s$^{-1}$ \\
\midrule
                                             Systemic redshift change &                   $\Delta v$$^{n}$ &           $-1$\,$\pm$\,$8$ &          $6^{+9}_{-13}$ &                                          km\,s$^{-1}$ \\
\midrule
\multirow{14}{*}{\makecell{\citet{Puccetti14} \\ contaminants}$^{o}$} &          log\,$kT_{{\tt apec}\,1}$ &     $-0.17$\,$\pm$\,$0.02$ &  $-0.17$\,$\pm$\,$0.02$ &                                                   keV \\
                                                                      &           log\,$N_{{\tt apec}\,1}$ &    $-4.69^{+0.06}_{-0.07}$ & $-4.68^{+0.06}_{-0.09}$ & $\frac{10^{-14}}{4\pi [D_A(1+z) ]^2} \int n_e n_H dV$ \\
                                                                      &         log\,$N_{\rm H,\,cont\,2}$ &      $0.30$\,$\pm$\,$0.10$ &  $0.32^{+0.08}_{-0.12}$ &                                             cm$^{-2}$ \\
                                                                      &          log\,$kT_{{\tt apec}\,2}$ &    $-0.02^{+0.07}_{-0.10}$ & $-0.07^{+0.08}_{-0.10}$ &                                                   keV \\
                                                                      &           log\,$N_{{\tt apec}\,2}$ &     $-3.25$\,$\pm$\,$0.18$ & $-3.25^{+0.17}_{-0.19}$ & $\frac{10^{-14}}{4\pi [D_A(1+z) ]^2} \int n_e n_H dV$ \\
                                                                      &         log\,$N_{\rm H,\,cont\,3}$ &     $0.89^{+0.06}_{-0.14}$ &   $1.03$\,$\pm$\,$0.16$ &                                             cm$^{-2}$ \\
                                                                      &          log\,$kT_{{\tt apec}\,3}$ &     $1.00^{+0.05}_{-0.02}$ &   $0.36$\,$\pm$\,$0.08$ &                                                   keV \\
                                                                      &           log\,$N_{{\tt apec}\,3}$ &    $-3.04^{+0.04}_{-0.05}$ & $-3.15^{+0.11}_{-0.12}$ & $\frac{10^{-14}}{4\pi [D_A(1+z) ]^2} \int n_e n_H dV$ \\
                                                                      &         log\,$N_{\rm H,\,cont\,4}$ &     $0.09^{+0.04}_{-0.05}$ &  $0.09^{+0.04}_{-0.05}$ &                                             cm$^{-2}$ \\
                                                                      &             $\Gamma_{\rm cont\,4}$ &      $2.18$\,$\pm$\,$0.01$ &  $2.22^{+0.01}_{-0.02}$ &                                                    -- \\
                                                                      &             log\,$N_{\rm cont\,4}$ &    $-2.86^{+0.00}_{-0.01}$ & $-3.09^{+0.03}_{-0.05}$ &                ph\,keV\,cm$^{-2}$\,s$^{-1}$ at 1\,keV \\
                                                                      &         log\,$N_{\rm H,\,cont\,5}$ &    $-0.09^{+0.05}_{-0.06}$ & $-0.09^{+0.05}_{-0.07}$ &                                             cm$^{-2}$ \\
                                                                      &          log\,$kT_{{\tt apec}\,5}$ &    $-0.77^{+0.02}_{-0.03}$ & $-0.77^{+0.02}_{-0.03}$ &                                                   keV \\
                                                                      &           log\,$N_{{\tt apec}\,5}$ &    $-1.72^{+0.11}_{-0.14}$ & $-1.70^{+0.11}_{-0.17}$ & $\frac{10^{-14}}{4\pi [D_A(1+z) ]^2} \int n_e n_H dV$ \\
\midrule
                                             \multirow{3}{*}{Fe\,XXV} &            E$_{\rm Fe\,XXV}$$^{p}$ &  $6.659^{+0.004}_{-0.003}$ & $6.662$\,$\pm$\,$0.004$ &                                                   keV \\
                                                                      &         FWHM$_{\rm Fe\,XXV}$$^{q}$ &       $4505^{+320}_{-379}$ &    $4486^{+333}_{-393}$ &                                          km\,s$^{-1}$ \\
                                                                      &       log\,$N_{\rm Fe\,XXV}$$^{r}$ &    $-5.10^{+0.03}_{-0.04}$ & $-5.10^{+0.03}_{-0.05}$ &                               ph\,cm$^{-2}$\,s$^{-1}$ \\
\midrule
                                       \multirow{2}{*}{Ni\,K$\alpha$} & log\,$L_{{\rm Ni\,K}\alpha}$$^{s}$ &     $37.75$\,$\pm$\,$0.07$ & $37.72^{+0.07}_{-0.11}$ &                                         erg\,s$^{-1}$ \\
                                                                      & log\,$Z_{{\rm Ni\,K}\alpha}$$^{t}$ &        $0.6^{+0.6}_{-0.3}$ &     $0.5^{+0.5}_{-0.3}$ &                                           Z$_{\odot}$ \\
\midrule
                                                        Fit statistic &    \texttt{cstat}\,/\,d.o.f.$^{u}$ &           3026.68\,/\,2405 &        2730.29\,/\,2400 &                                                    -- \\
\bottomrule
\end{tabular*}
\vspace{5pt}

\begin{minipage}{\textwidth}
\small
\textbf{Notes.} $^{a}$Photon index, $^{b}$exponential cut-off energy (fixed) and $^{c}$intrinsic 2\,--\,10\,keV luminosity of the primary X-ray coronal emission. $^{d}$Eddington ratio estimated with the Compton-thick X-ray-to-bolometric luminosity correction of \citet{Brightman17} and the black hole mass of \citet{Greenhill97}. $^{e}$line-of-sight column density, $^{f}$covering factor, inclination angle (fixed), $^{h}$Gaussian broadening FWHM for the primary coupled obscurer. $^{i}$Relative scaling of the reprocessed emission, $^{j}$line-of-sight column density, $^{k}$covering factor, $^{l}$inclination angle, $^{m}$Gaussian broadening FWHM for the secondary reprocessing component. $^{n}$Velocity offset relative to the systemic host galaxy redshift. $^{o}$Modelled sources of off-nuclear contaminating X-ray flux constrained by \citet{Puccetti14} -- see §\ref{sec:offnuc} for more information. $^{p}$Centroid energy, $^{q}$FWHM and $^{r}$logarithmic normalisation of the ionised Fe\,XXV broad Gaussian component included as a nuisance parameter (c.f. §\ref{sec:fexxv}). $^{s}$Luminosity and $^{t}$corresponding abundance of Nickel inferred in solar units for the additional Nickel emission included as a nuisance parameter (c.f. §\ref{subsec:nika}). $^{u}$C-statistic divided by the number of degrees-of-freedom in the fit.
\end{minipage}
\end{table}


\end{appendix}
\end{document}